\documentclass[]{aastex631}

\usepackage{amsmath}
\usepackage{graphicx}
\usepackage{acronym}
\usepackage{appendix}
\usepackage{diagbox}
\usepackage[normalem]{ulem}
\usepackage{booktabs}
\usepackage{tabu}
\usepackage[T1]{fontenc}
\usepackage{pgfgantt}
\acrodef{lvk}[LVK]{LIGO-Virgo-KAGRA}
\acrodef{bbh}[BBH]{Binary black hole}
\acrodef{bns}[BNS]{binary neutron star}
\acrodef{sfr}[SFR]{star formation rate}
\acrodef{gw}[GW]{gravitational-wave}
\acrodef{cbc}[CBC]{compact binary coalescence}
\acrodef{los}[LOS]{line-of-sight}
\acrodef{pe}[PE]{parameter estimation}
\acrodef{em}[EM]{electromagnetic}
\acrodef{o5}[O5]{fifth observing run}
\acrodef{snr}[SNR]{signal-to-noise ratio}

\begin{document}

\title{Using gravitational wave dark sirens to choose between host galaxy weighting models}


\author{Zhuotao Li}
\email{z.li.13@research.gla.ac.uk}
\affiliation{SUPA, University of Glasgow \\
Glasgow, G12 8QQ, United Kingdom}

\author{Rachel Gray}
\affiliation{SUPA, University of Glasgow \\
Glasgow, G12 8QQ, United Kingdom}

\author{Ik Siong Heng}
\affiliation{SUPA, University of Glasgow \\
Glasgow, G12 8QQ, United Kingdom}

\begin{abstract}
\ac{bbh} mergers, an important source of \acp{gw}, are assumed to be hosted in galaxies. The probability of a galaxy to host a BBH is related to its properties, for example stellar mass and star formation rate. These properties can be estimated from observables, such as the luminosity in certain observation bands. We refer to this description of host galaxy properties as host galaxy weighting models. However, the host galaxy weighting model for \ac{bbh}s has yet to be accurately determined. Population synthesis has provided a variety of host galaxy weighting models. Here we investigate whether it is possible to distinguish different host galaxy weighting models using a data driven approach. We use the GW cosmology tool gwcosmo with a simulated \ac{lvk} \ac{o5}-like observing scenario. We also construct a mock spectroscopic galaxy catalogue from MICECATv2. Our analysis compares the Bayes factors among three simple luminosity weighting models, r-band, g-band, and uniform weighting. The Bayes factors among different host galaxy weighting models show a minor preference for the true model for a $\sim$200-detection case, and a decisive preference for the true model over the uniform model for a $\sim$1000-detection case, which is strongly driven by a small number of well-localised events.

\end{abstract}



\section{Introduction}\label{Sec:1}

\ac{bbh}s account for the majority of \acp{gw} detected to date. They are generally assumed to be hosted in galaxies. However, it is believed that some galaxies are more likely to host \ac{bbh}s than others based on their galaxy properties, e.g. massive galaxies are believed to be more likely to host \ac{bbh}s compared to low-mass galaxies in general. 

So far, there have been many studies that have come to the conclusion that certain types of galaxies should have a higher likelihood to host \ac{bbh}s. These studies estimate the \ac{bbh} merger rate of galaxies by simulating \ac{bbh} formation channels (isolated or dynamical) in the galaxies, and the \ac{bbh} formation channels will be related to the host galaxy properties. (We refer to this description of the distribution of galaxy properties for host galaxies as "host galaxy weighting model" in the following.) For example, \citet{bbhformRauf2023} simulated the binary population synthesis in the galaxies from a mock universe, and found that metal-rich and massive galaxies with \ac{sfr} higher than $1\mathrm{M_\odot/yr^{-1}}$ had higher \ac{bbh} merger rate. Another example, \citet{bbhform3Neijssel2019}, focused more on the impact of metallicity-specific star formation rate on the \ac{cbc} merger rate of the host galaxies. Other attempts to constrain the relationship between merger rate and galaxy intrinsic properties include \citet{bbhform7_Santoliquido2021,bbhform6_Broekgaarden2022,bbhformation_Srinivasan2023,bbhform5_Chruślińska2024} etc., all based on similar ideas that we can use \ac{bbh} formation channels to estimate the host galaxy weighting models. Similar studies exist for \ac{bns} as well, for example \citet{bns_formation1,bns_formation2,bns_formation3,bns_formation4,bns_formation5,BNS_DH}, etc.

Further more, the merger rate of galaxies can be linked to the observables as indicators for galaxy properties. These intrinsic galaxy properties, e.g. stellar mass, \ac{sfr}, metallicity, etc., determine the observables including luminosity of certain bands, $H_\alpha$ lines, colour, etc. There have been many studies on galaxy property indicators, e.g.  \citet{sfr_indi,stellar_mass_indi} discuss the indicators of \ac{sfr} and stellar mass. However, the relationship between indicators and galaxy properties is a very complicated issue, so for simplicity it is easier to first consider the direct relationship between merger rate and observables without galaxy properties as intermediate. For example, \citet{Suvodip1} directly proposed to use the emission lines of galaxies to constrain the delay time function of \ac{bbh}s. This is what we will be doing in this paper.

All of the studies mentioned above try to constrain the host galaxy weighting models based on \ac{bbh} formation channels. However, we can also approach the problem of constraining host galaxy weighting model in a data-driven manner, i.e. starting from \ac{bbh} \ac{gw} data and galaxy catalogue data. Currently, there has been an attempt by \citet{rate_infer} to constrain the host galaxy weighting model from the redshift evolution of the BBH merger rate using data from \citet{O3}. To be more specific, \citet{rate_infer} used universe simulation to test what kind of host galaxy weighting model is needed in order to have a redshift distribution of \ac{bbh} events obtained from data under the assumption that the true weighting model is a combination of stellar mass and star formation rate. There are also attempts from \citet{Suvodip2,Suvodip3} to explore the difference between the overall matter distribution and \ac{bbh} distribution under different host galaxy weighting models in order to shed light on this problem from a large-scale structure perspective. Compared to these studies, however, we would like to focus more on how the localisation of individual \ac{bbh} \ac{gw} events with a galaxy catalogue can help to constrain the host galaxy weighting model.

Understanding the host galaxy weighting model is also very useful for many researches in \ac{gw} astronomy, for instance \ac{gw} cosmology and \ac{bbh} population studies. This is especially true for the dark siren analysis in \ac{gw} cosmology, which takes the host galaxy weighting model as input. The dark siren analysis is based on the assumption of \ac{bbh}s are hosted in galaxies, so it combines the luminosity distance information of a \ac{bbh} \ac{gw} event with redshift information from a galaxy catalogue, to infer the Hubble constant and other cosmological parameters \citep{H0inf_schutz1986,H0methods}. Here, the host galaxy weighting model plays an important role as it determines which galaxies are more likely to host the \ac{bbh}s and mishandling it could significantly bias the result \citep{weight_model_bias2,weight_model_bias1}. On the other hand, this also means the test of host galaxy weighting models can also be done using \ac{gw} cosmology analysis tools. At this point, this is not very feasible given the current detector sensitivity and galaxy catalogue, as is mentioned above that the results of different host galaxy weighting models did not differ enough in \citet{O3}. However, it could be possible with future \ac{o5} \ac{lvk} detector network and better galaxy catalogues which may include new sky surveys like DESI and EUCLID \citep{euclid,O5,DESI,whitepaper2024}.

So in this paper, we carry out a mock data analysis to perform Bayesian model comparisons for three host galaxy weighting models with the \ac{gw} cosmology analysis tool gwcosmo and the mock catalogue MICECATv2 (See Sec.\ref{Sec:3.1}). We first generate mock \ac{bbh} populations and inject them into a subset of MICECATv2 based on a ``true'' host galaxy weighting model. Then the mock \ac{gw} events of the \ac{bbh} populations are generated by the fisher matrix tool GWFish (see Sec.\ref{Sec:3.2}). In this case, we use DESI r-band luminosity weighting for injection and compared it to DESI g-band luminosity weighting and uniform weighting. Next, we use gwcosmo to calculate the Bayes factors among these three host galaxy weighting models and test if it is possible to distinguish them. To simplify the test, we fixed the cosmology and the population parameters in gwcosmo but tried both cases with the rate evolution model fixed or unfixed (see Eq.\ref{Eq:gwcosmo} in Sec.\ref{Sec:2} for definition of rate evolution model). In the end, our results show that under our setup, $\sim$200 detections could show minor preference over the true model while for $\sim$1000 detections the preference over the true model can be decisive against uniform weighting. The results are also decided by small number of well-localised events.

The structure of our work is organised as follows. In Sec.\ref{Sec:2} we talk about the theory behind gwcosmo and how to use it to perform host galaxy weighting model comparison. In Sec.\ref{Sec:3}, the mock data setup in this paper will be discussed and the mock galaxy catalogue MICECATv2 as well as the GWFish pipeline will be introduced in Sec.\ref{Sec:3.1} and Sec.\ref{Sec:3.2} respectively. In Sec.\ref{Sec:4} we will show the results of our analysis. The conclusion and a discussion will be given in Sect.\ref{Sec:5}

\section{Methodology}\label{Sec:2}
As is mentioned in Sec.\ref{Sec:1}, we modified the \ac{gw} cosmology pipeline, gwcosmo 2.0, to compare the host galaxy weighting models in this paper. gwcosmo is a pipeline developed by \citet{gwcosmo3,gwcosmo2,gwcosmo1} to estimate cosmological and \ac{gw} population parameters using gravitational-wave observations, both bright sirens and dark sirens. In gwcosmo, the dark siren method with galaxy catalogues combines the redshift information of the galaxies from the catalogue with the parameter posteriors of \ac{gw} events to infer cosmological and \ac{gw} population parameters. The analysis is done by Bayesian analysis with the mathematical expression of cosmological parameter posteriors as:
\begin{equation}\label{Eq:gwcosmo}
\begin{aligned}
 p\left(\Lambda \mid\left\{x_{\mathrm{GW}}\right\},\left\{D_{\mathrm{GW}}\right\}, I\right) \propto &p(\Lambda \mid I) p\left(N_{\mathrm{det}} \mid \Lambda, I\right) \mathrm{S_{election}}  \\ & \times \prod_i^{N_{\mathrm{det}}}\left[\iint \sum_j^{N_{\mathrm{pix}}} p\left(x_{\mathrm{GW} i} \mid \Omega_j, z, \theta^{\prime}, \Lambda, I\right) p\left(\theta^{\prime} \mid \Lambda, I\right) p\left(z \mid \Omega_j, \Lambda, I\right) d \theta^{\prime} d z\right] 
\end{aligned}
\end{equation}
Here $\Lambda=\{\Lambda_{\mathrm{cosmo}},\Lambda_{\mathrm{population}},\Lambda_{\mathrm{rate}}\}$ means the hyper-parameters of interest including cosmological parameters $\Lambda_{\mathrm{cosmo}}$, \ac{gw} source population parameters $\Lambda_{\mathrm{population}}$, and the parameters $\Lambda_{\mathrm{rate}}$ that describe rate evolution models, defined as how the rate of the \ac{gw} source evolves with redshift. In addition, $x_{\mathrm{GW}}$ means the gravitational wave data, $D_{\mathrm{GW}}, N_{det}$ mean the detection of a GW event and the number of detections, $\Omega, z, \theta$ mean the sky location, redshift, and other parameters of the GW source, and $I$ represents all the other background information. In addition, since gwcosmo divides the sky into pixels to calculate catalogue incompleteness, here $j$ represents individual pixels and $N_\text{pix}$ means the total number of healpix pixels. Eq. \ref{Eq:gwcosmo} basically states that the posterior on cosmological parameters depends the prior of the parameter $p(\Lambda \mid I)$, the detection probability $p(N_{det}\mid \Lambda,I)$, the selection bias correction $\mathrm{S_{election}}$(the fact that not all events can be seen), the GW likelihood $p\left(x_{\mathrm{GW} i} \mid \Omega_j, z, \theta^{\prime}, \Lambda, I\right)$ and the line-of-sight (LOS) redshift prior $p\left(z \mid \Omega_j, \Lambda, I\right)$ mentioned in Sec.\ref{Sec:1}  which will contain the information of host galaxy weighting model as is shown in the following.

For the purpose of this paper, the \ac{gw} likelihood, which can be obtained from the GW \ac{pe} posteriors, and \ac{los} priors are of the main interest. \citet{gwcosmo1} shows that \ac{los} prior, i.e. the redshift prior brought in by the galaxy catalogue, can be divided into host galaxy in the catalogue term and out of catalogue term. But in this paper, we assume the catalogue to be complete, i.e. we can see every galaxy in the universe. Thus, we only consider the in-catalogue part which has an expression as follows:
\begin{equation}\label{Eq:los}
\begin{aligned} p\left(z \mid \Omega_i, \Lambda, s, I\right)  \propto p(s\mid z,\Lambda_{\mathrm{rate}},I)\left [ \frac{1}{N_{\mathrm{gal}}\left(\Omega_i\right)} \sum_k^{N_{\mathrm{gal}}\left(\Omega_i\right)} p\left(z \mid \hat{z}_k\right) p\left(s \mid z, M\left(z, \hat{m}_k, \Lambda_{\mathrm{cosmo}}\right), I\right)\right ], \end{aligned}
\end{equation}
where $s$, hidden inside $I$ in Eq.\ref{Eq:gwcosmo}, represents the GW source hosted by the galaxy, $M$ and $m$ are the absolute and apparent magnitudes of the galaxy. Eq.\ref{Eq:los} basically sums over the redshift distribution of all galaxies within one pixel along the line-of-sight, weighs them using $p(s \mid z, M(z, \hat{m}_k, \Lambda_{\mathrm{cosmo}}), I)$ as their likelihood of hosting GW sources, and multiplies it with the rate evolution model $R(z)=p(s\mid z,\Lambda_{\mathrm{rate}},I)$. The term $p(s \mid z, M(z, \hat{m}_k, \Lambda_{\mathrm{cosmo}}), I)$ is exactly the mathematical representation of the host galaxy weighting model, the main focus of this paper. During the GWTC-3 analysis \citet{O3}, K band and $B_J$ band luminosities for the GLADE+ galaxies have been used as the weights, assuming the distribution in Eq.\ref{Eq:lumin_weighted}. See \citet{GLADE,GLADE+} for details about the GLADE+ catalogue.

\begin{equation}\label{Eq:lumin_weighted}
p(s \mid z, M(z, \hat{m}_k, \Lambda)) \propto L(M(z,\hat{m}_k,\Lambda_{\mathrm{cosmo}}),I),
\end{equation}
where $L(M(z,\hat{m}_k,\Lambda_{\mathrm{cosmo}}),I)$ here is the luminosity of the galaxy as a function of K-band apparent magnitude $M$. In addition to K-band and $B_J$ band weighting, there is also the uniform weighting model where each galaxy is viewed to have the same probability of hosting \ac{bbh}s. To compare these different host galaxy weighting models, we calculate the Bayes factors, i.e. compare the evidence of each model. According to Bayesian analysis, the evidence should be calculated by marginalizing the likelihood and prior over the parameter space. In our paper, we fixed the cosmology and \ac{bbh} population parameters $\Lambda_{\mathrm{cosmo}}$ and $\Lambda_{\mathrm{population}}$ for simplicity so the parameter space will be a delta function. And for the rate model $p(s\mid z,\Lambda_{\mathrm{rate}},I)$ we have used both a simple Powerlaw model with index $\gamma$ and Madau-Dickinson rate evolution model. The expression for the Powerlaw model is simply:
\begin{equation}\label{Eq:powerlaw}
p(s\mid z,\gamma,I) \propto (1+z)^\gamma,
\end{equation}
and the expression for Madau-Dickinson rate evolution model according to \citet{madau2014cosmic} is:
\begin{equation}\label{Eq:madau}
    P_{\mathrm{Madau-Dickinson}}(s|z,\gamma,k,z_p,I)=[1+(1+z_p)^{-\gamma-k}]\frac{(1+z)^{\gamma}}{1+[(1+z)/(1+z_p)]^{\gamma+k}}
\end{equation}
For the simple Powerlaw model, we test both cases where $\gamma$ is fixed at 0, i.e. no rate evolution, or unfixed (see Sec.\ref{Sec:4} for more details). This is because the rate evolution of \ac{bbh}s is determined by the true host galaxy weighting model and can provide information about this true model. However, in this paper we only concern the information from individual events, so we unfix the rate evolution model to remove the extra information from rate evolution. This is also more realistic since in real life we do not have the true host galaxy weighting model. So the final expression of evidence is:
\begin{equation}\label{Eq:evidence}
\begin{aligned}
\mathcal{Z}_i & \propto \int p\left(\Lambda \mid\left\{x_{\mathrm{GW}}\right\},\left\{D_{\mathrm{GW}}\right\},w_i, I\right) p(\Lambda|I) \mathrm{d}\Lambda
\\ & \propto  \int p\left(\gamma \mid\left\{x_{\mathrm{GW}}\right\},\left\{D_{\mathrm{GW}}\right\},w_i,\Lambda^\prime, I\right)p(\gamma\mid I)\mathrm{d}\gamma,
\end{aligned}
\end{equation}
where $\mathcal{Z}_i$ is the evidence for the i-th model, $w_i$ represents different weighting models, $p(\gamma\mid I)$ is assumed to be uniform, and $\Lambda^\prime = \{\Lambda_{\mathrm{cosmo}},\Lambda_{\mathrm{population}}\}$. With the expression of evidence, the Bayes factor between host galaxy weighting models is just the ratio between their evidence, $B_{\frac{i}{j}}=\mathcal{Z}_i/\mathcal{Z}_j$.

\begin{figure}[htbp] 
    \centering
    \includegraphics[width=0.9\textwidth]{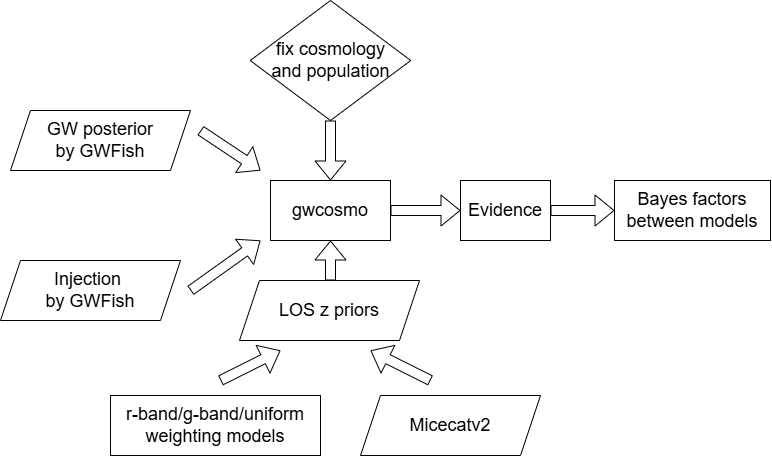}
    \caption{This is the flow chart of this paper, showing how we used gwcosmo to calculate the evidence of different host galaxy weighting models. To output the evidence, gwcosmo requires input of GWFish-generated mock \ac{gw} events, selection effect, and the \ac{los} redshift prior generated by mock galaxy catalogue MICECATv2 and different host galaxy weighting models. And by comparing the evidence we have the Bayes factors}
    \label{fig:flowchart}
\end{figure}

We apply the method above on GWFish processed mock \ac{bbh} \ac{gw} events which are injected into mock galaxy catalogue MICECATv2 with cosmology and \ac{bbh} population model fixed as is shown in flowchart Fig.\ref{fig:flowchart}. The setup of the mock data as well as host galaxy weighting model choices will be discussed in Sec.\ref{Sec:3}.

\section{Mock data}\label{Sec:3}

\subsection{Mock catalogue: MICECATv2}\label{Sec:3.1}
To generate the \ac{los} redshift prior in Eq.\ref{Eq:los}, the mock galaxy catalogue MICECATv2 (MICE-Grand Challenge Galaxy and Halo Light-cone Catalog) is used in a complete catalogue case. It is built on the MICE cosmology simulation and available on COSMOHUB\footnote{https://cosmohub.pic.es}, a distribution centre of a range of mock or real galaxy catalogues\citep{cosmohub2,cosmohub1}.

The MICE Grand Challenge simulation is an N-body simulation of about 70 billion dark matter particles in a flat $\Lambda\mathrm{CDM}$ universe with $H_0=70 \mathrm{km/s/Mpc}$ and $\Omega_m=0.25$, which will be the cosmology setup in this paper as well \citep{micecatv2_1,micecatv2_3,micecatv2_2}. The dark matter particles in MICE simulations form halos due to gravity and MICECATv2 used a halo occupation distribution (HOD) recipe and the subhalo abundance matching (SHAM) techniques to generate galaxies from these dark matter halos. The results were also calibrated to observations. See \citet{micecatv2_5,micecatv2_4} for details about the generation of the catalogue.

MICECATv2 offers the absolute and apparent magnitudes of 500 million galaxies in the observation bands of the DESI (Dark Energy Spectroscopic Instrument) telescope, calibrated by r-band luminosity Schechter function based on real local universe observations \citep{DESI}. Covering one octant of the sky, it claims to be complete for DES-like surveys out to redshift $z=1.4$ (for i-band apparent magnitude smaller than 24) with low mass galaxies from dark matter halos with as low as 10 dark matter particles.

\begin{figure}[htbp] 
    \centering
    \includegraphics[width=0.9\textwidth]{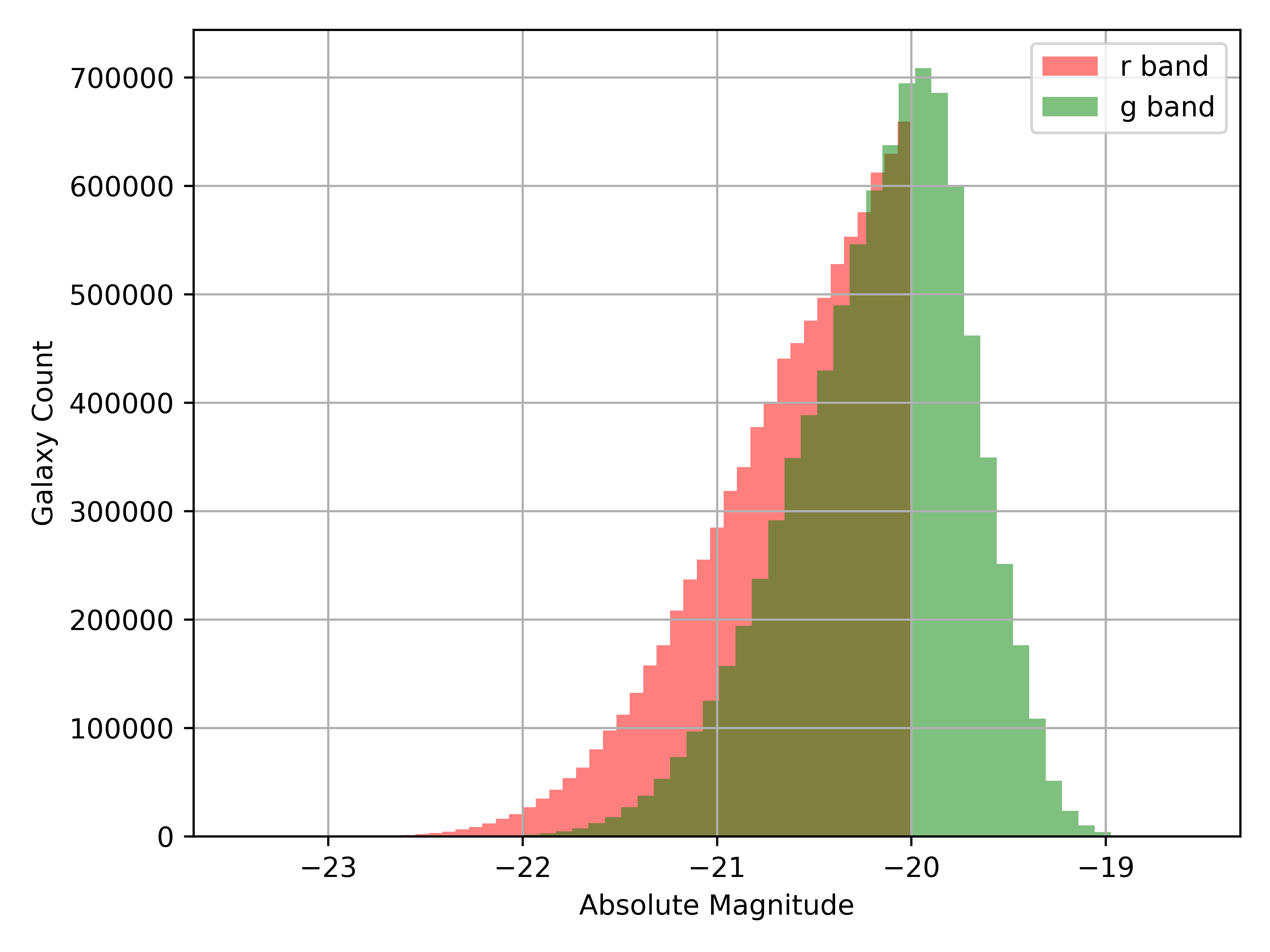}
    \caption{This is the distribution of galaxies over r-band and g-band absolute magnitude used in this paper. These galaxies are randomly sampled 10\% from the subset of MICECATv2 with r-band absolute magnitude cut at -20. }
    \label{fig:sliced_catalog}
\end{figure}

Given the vast number of galaxies in MICECATv2, it is necessary to draw samples of galaxies to reduce computational cost. In this paper we first slice MICECATv2 at r-band absolute magnitude -20, ignoring all the galaxies fainter than this threshold. We apply this magnitude cut also because we would like to make our mock galaxy catalogue volume limited, i.e. being a complete catalogue given this r-band absolute magnitude threshold for the whole redshift range. By applying this cut we assume only the brightest galaxies are able to host \ac{bbh}s for this initial research, similar to \citet{brightsubset}. For the comparison results of a higher cut on r-band absolute magnitude (i.e., including more faint galaxies), please check Appendix \ref{Sec:appendix}.
We then randomly draw 10\% from galaxies that passed the -20 r-band threshold and assume this sliced catalogue to be complete to $z=1.4$ with around 9 million galaxies used in total and hundreds of galaxies in one pixel on the nside=128 healpix skymap used in this paper. It should be noted that this artificial low density of galaxies will make our results more optimistic since fewer galaxies means easier to locate the host galaxy. 
The distribution of galaxies used in this paper over r-band and g-band luminosity is shown in Fig.\ref{fig:sliced_catalog}.

For the redshift uncertainty of the galaxies in our mock catalogue, we assume spectroscopic redshift. For each galaxy, the redshift has a Gaussian distribution centring at their true redshift with the standard deviation being:
\begin{equation}\label{Eq:redshiftuncertainty}
\sigma_i = 0.001\times(1+z_i), 
\end{equation}
where $z_i$ and $\sigma_i$ are the redshift and redshift uncertainty of the i-th galaxy. 

\begin{figure}[htbp] 
    \centering
    \includegraphics[width=0.9\textwidth]{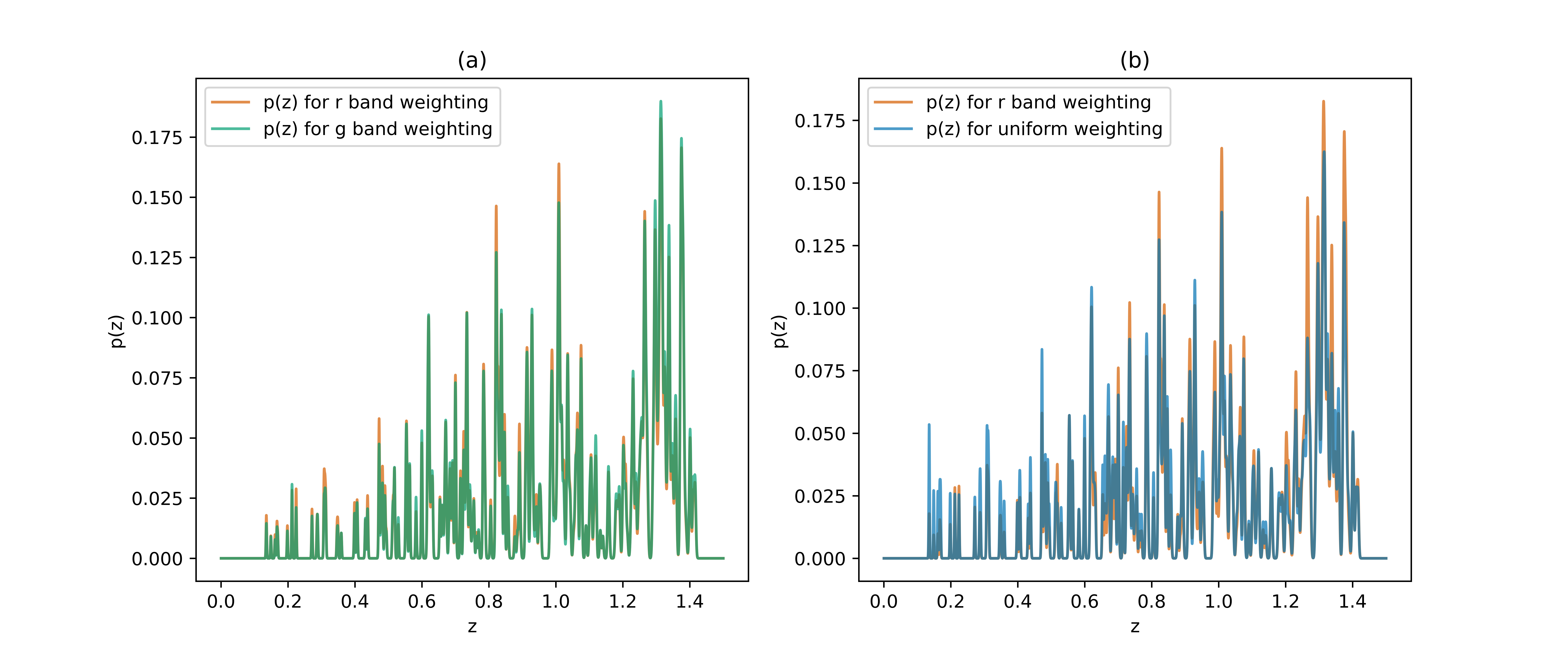}
    \caption{This is the comparison of the \ac{los} redshift prior between different weighting models of an example pixel. The left panel (a) compares r-band weighting and uniform weighting while the right panel (b) compares r-band weighting and g-band weighting. The resolution of the pixel is 128. Note that the uniform weighting has a varying $p(z)$ because: first the width of the Gaussian distribution is increasing with redshift according to Eq.\ref{Eq:redshiftuncertainty}, second the galaxies tend to cluster together based on the dark matter halo.}
    \label{fig:example_pixel_runi}
\end{figure}

Since MICECATv2 is calibrated to the DESI r-band, the host galaxy weighting models we decided to test in the end are the luminosity weighting models of DESI observation bands including r-band luminosity weighting, g-band luminosity, and uniform weighting. We injected mock \ac{gw} events into MICECATv2 galaxies using r-band weighting and calculated the \ac{los} redshift priors of these three models for the sliced catalogue respectively. The \ac{los} redshift prior comparison between r-band weighting, g-band weighting and uniform weighting of a random example pixel on a nside=128 healpix skymap is shown in Fig.\ref{fig:example_pixel_runi}, and it is clear that the redshift prior from the three models is different and this is where the constraining power of this approach comes from. It should also be noted that r-band weighting differs from uniform weighting more than g-band weighting, so we would expect that the method distinguish r-band weighting and uniform weighting better than r-band and g-band weighting. This also means our results for r-band versus g-band would likely to be more pessimistic compared to other optical bands since they are quite close to each other.

\subsection{Mock \ac{bbh} \ac{gw} events}\label{Sec:3.2}
Since we use dark siren analysis method to test different host galaxy weighting models, it is necessary to generate the GW likelihood $p\left(x_{\mathrm{GW} i} \mid \Omega_j, z, \theta^{\prime}, \Lambda, I\right)$ in Eq.\ref{Eq:gwcosmo}. This requires simulating detection results and \ac{pe} posteriors of mock BBH populations. In this paper, we simulate the \ac{pe} posterior outputs for detected \ac{bbh} events with fisher matrix analysis. We assume an \ac{o5} sensitivity three detector network : LIGO-Hanford, LIGO-Livingston, and Virgo for this initial paper\citep{avirgo,aligo,O5,effler2023roadmap,whitepaper2024}. The mock \ac{bbh} populations and \ac{pe} results are generated in the following manner:

We first generate the source parameters of the \ac{bbh} population. The masses of the \ac{bbh}s are generated in the source frame using the BBH Powerlaw+Peak mass model the same with GWTC-3 setup \citet{O3}. Other parameters within consideration here follow a uniform distribution within their prior range, namely $\{\theta_{JN},\Psi,\phi,t_c\}$. The parameters for the Powerlaw+Peak model and the prior ranges for $\{\theta_{JN},\Psi,\phi,t_c\}$ are shown in Tab.\ref{tab:mass_prior}. In terms of the rate evolution model $p(s|z,\gamma,I)$, we choose $\gamma=0$ in the Powerlaw mode in Eq.\ref{Eq:powerlaw}. 

Once generated, we injected these mock \ac{bbh} events into MICECATv2 galaxies based on the r-band luminosity weighting of each galaxy. The \ac{bbh} mergers follow the redshift and sky location of their host galaxies. 

The optimal \ac{snr} of these mock events will be calculated, and we apply the \ac{snr} threshold 18 to the optimal \ac{snr}s. The events with \ac{snr} passed this threshold will be considered as detected proceed for mock \ac{pe}. This \ac{snr} threshold 18 is higher than normally used 11 because GWFish performs better at higher \ac{snr}, and higher \ac{snr} means better luminosity distance uncertainty and localisation.

The likelihood of the detected events will then be calculated using Fisher matrices with the tool GWFish \citep{gwfish}. Fisher matrices simulated \ac{pe} results are cheap to generate and thus suitable for this paper which is expected to require a large number of GW events to tell the differences among host galaxy weighting models. 

The Fisher matrix method is based on the Gaussian approximation of the likelihood function which is valid for high \ac{snr} signals:
\begin{equation} \label{Eq:Fisher_likelihood}
\mathcal{L} \propto \exp \left(-\frac{1}{2} \Delta \theta^i \mathcal{F}_{i j} \Delta \theta^j\right),
\end{equation}
where $\Delta \theta = \theta - \bar{\theta}$ is the error from the true value of the parameter vector $\bar{\theta}$. Here $\mathcal{F}_{i j}$ are the Fisher matrices which are calculated from the derivative of a waveform with respect to the parameters $\theta$ given some detector noise spectral density. It is also the inverse of the covariance matrix. We refer to \citet{gwfish} again for detailed expression of the fisher matrices. The waveform model we used for derivative calculation is IMRPhenomXPHM \citep{waveform}. From the input parameters of IMRPhenomXPHM, we consider the following parameters to be included in $\theta$, i.e., the parameters that are used to calculate the derivative:
\begin{equation} \label{Eq:Fisher_para}
\theta=\{m_{1,d},m_{2,d},d_L,\theta_{JN},\delta,\alpha,\Psi,\phi,t_c\},
\end{equation}
where $m_{1,d},m_{2,d}$ are the detector-frame primary and secondary masses, $d_L$ is the luminosity distance, $\delta$ is the declination, $\alpha$ is the right ascension, and other parameters have already been mentioned above. 

We then calculate the posterior $p(\Omega,z,\theta \mid x_{GW_i},\Lambda,I)$ and draw posterior samples from it. We first apply a prior $p(\Omega,z,\theta\mid\Lambda,I)$ on the likelihood $p\left(x_{\mathrm{GW} i} \mid \Omega_j, z, \theta^{\prime}, \Lambda, I\right)$ to calculate $p(\Omega,z,\theta \mid x_{GW_i},\Lambda,I)$. The mass and luminosity priors are the same with the GWTC-3 $H_0$ inference paper \citet{O3} since gwcosmo 2.0 will automatically remove these priors from posterior samples to compute likelihood. These priors are a uniform distribution for the detector frame mass and a $d_L^2$ prior for the luminosity distance: $p(d_L(z,\Lambda))\propto d_L^2$. The rest of the parameters are not of concern here. The posterior samples will then be drawn from $p(\Omega,z,\theta \mid x_{GW_i},\Lambda,I)$ using rejection sampling, during which samples of with $m_1<m_2$ will be thrown away to get the correct primary mass and secondary mass distribution. 

\begin{figure}[htbp] 
    \centering
    \includegraphics[width=0.9\textwidth]{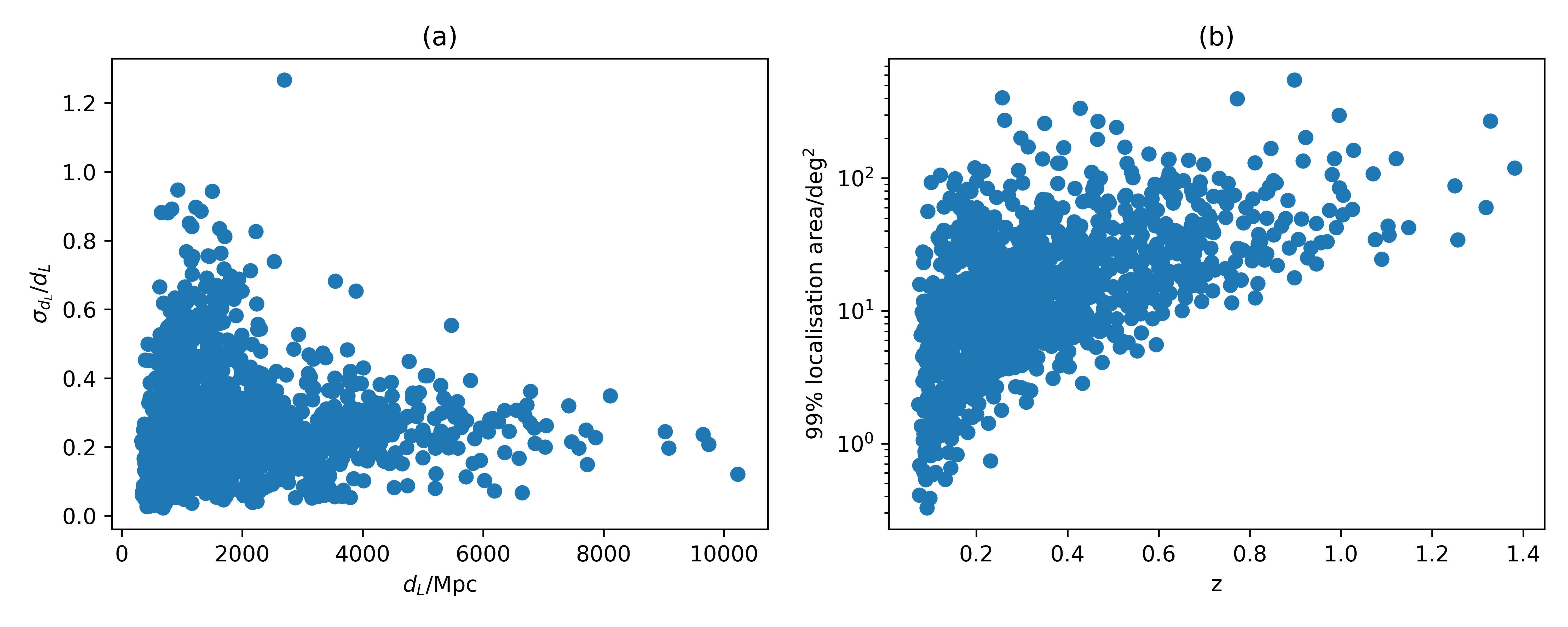}
    \caption{Here are two plots showing the 1 $\sigma$ uncertainty distribution estimated by GWFish. The left panel (a) shows the distribution of uncertainties of luminosity distance in ratio while the right panel (b) shows the distribution of 99\% sky localisation area against redshift. This example population consists of 214 mock \ac{bbh} detection.}
    \label{fig:error}
\end{figure}

To illustrate the uncertainty of of the posterior under the detector network setup we used in this paper, we show the distribution of $\sigma_{d_L}/d_L$ for luminosity distance and 99\% of the sky localisation area in Fig.\ref{fig:error}. 

Finally, to correctly perform the inference using gwcosmo, it is necessary to calculate the selection effects in Eq.\ref{Eq:gwcosmo}.  The selection effect determines which part of source parameter space is detectable and the according bias from this. For gwcosmo, the input for selection effect calculation is a list of detectable (i.e. able to pass the \ac{snr} threshold) GW events generated in the detector frame to calculate selection effects. A detailed derivation of selection effect can be found in \cite{gwcosmo1}. To generate this list of detectable events in our paper, we first calculate the optimal SNR of a \ac{bbh} population which has a wide detector frame mass prior, uniform from 1 to 500 solar masses. The wide detector frame mass prior is applied because it covers large parameter space for source frame mass and redshift. We then apply the SNR threshold of 18 to the \ac{snr} of these event and record the events that can pass the threshold. These \ac{bbh}s follow a redshift prior uniform in comoving volume.



\begin{table}[]\label{tab:mass_prior}
\begin{tabular}{lll}
\hline
Parameters   & Description                                                               & Value\\ \hline \hline
$M_{min}$       & Minimum value for black hole source frame mass prior                      & 4.98  \\
$M_{max}$       & Maximum value for black hole source frame mass prior                      & 112.5 \\
$\alpha$        & Powerlaw index for the primary mass                                       & 3.78  \\
$\mu_g$        & Centre of the Gaussian component in the primary mass distribution         & 32.27 \\
$\sigma_g$     & Standard error of the Gaussian component in the primary mass distribution & 3.88  \\
$\lambda_{peak}$ & Fraction of the model in the Gaussian component                           & 0.03  \\
$\delta_m$     & Range of mass tapering on the lower end of the mass distribution          & 4.8   \\
$\beta$         & Powerlaw index for the secondary mass                                     & 0.81 
\\ 
\hline
$\theta_{JN}$ & Angle between the total angular momentum vector and the line of sight to the observer & $\mathrm{U}(-\pi,\pi)$
\\
$\Psi$ & Polarization angle&$\mathrm{U}(0,2\pi)$
\\
$\phi$& Phase&$\mathrm{U}(0,2\pi)$
\\
$t_c$& Geocentric time&$\mathrm{U}(0,t_{max})$
\\
\hline
$H_0$         & Hubble constant                                     & 70.0
\\
$\Omega_m$         &  Matter energy density                                & 0.25\\ \hline
\end{tabular}
\caption{Here are the parameters for the Powerlaw+Peak mass model, the uniform distributions of $\theta_{JN},\Psi,\phi,t_c$, and the cosmology parameters used to generate the \ac{bbh} population.}
\end{table}




\section{Analysis and results}\label{Sec:4}

 We first conducted our tests with three populations (Population 1,2,3) and fixed cosmological parameters. Each population has 5000 \ac{bbh} injections drawn from priors described in Sec.\ref{Sec:3.2}. From these injections, around 200 of them are detected. All different populations are drawn from the same mass model as mentioned in Tab.\ref{tab:mass_prior} and injected into MICECATv2 galaxies based on r-band weighting. We test multiple populations to quantify the variation that can arise from the random sample. For each population, we run our tests in two cases: one is fixing the rate evolution parameter $\gamma$ in Eq.\ref{Eq:powerlaw} to be 0 and the other one is unfixing $\gamma$. This is to show the influence of the \ac{bbh} rate evolution of each host galaxy weighting model in our analysis as we have explained in Sec.\ref{Sec:2}. We present the full-sky redshift priors for different host galaxy weighting models and the Bayes factors of individual events to further explain this fact. According to \citet{O3_ratespop}, the estimated \ac{bbh} merger rate is between 17.9 and 44 $\mathrm{Gpc}^{-3} \mathrm{yr}^{-1}$ increasing with redshift at a rate proportional to $(1+z)^\kappa$ with $\kappa=2.9$. Based on this \ac{bbh} rate and our detection efficiency, i.e. we can detect around 200 \ac{bbh}s out of 5000 \ac{bbh} injections, the $\sim$200~detection case would be an estimation of the \ac{o5} \ac{lvk} run lasting for a few months. 

 We also test two more cases with $\sim$1000~detection populations (Population 4,5) and $\gamma$ unfixed to show the effect of adding more \ac{gw} events. This number of detections represents a hypothetical scenario of \ac{o5} lasting for over a year given the estimated \ac{bbh} detection rate mentioned above. The \ac{bbh}s in the two $\sim$1000~detection populations are generated and injected into MICECATv2 galaxies in the exact same way with other populations (Population 1,2,3).

 Finally, we substitute the simple Powerlaw rate model described in Eq.\ref{Eq:powerlaw} with the more sophisticated Madau-Dickinson rate evolution model described in Eq.\ref{Eq:madau} to test if Madau-Dickinson model compensates the rate evolution information better. We apply the Madau-Dickinson model to the analysis of the three $\sim$200-detection populations (Population 1,2,3) and compare the Bayes factors results to the results of the simple Powerlaw model.

\subsection{Rate evolution model parameter $\gamma$ fixed case}\label{Sec:4.1}

\begin{table}[]
\begin{tabular}{|l|c|c|c|}
\hline
Bayes factors                 & \multicolumn{1}{l|}{r-band / g-band} & \multicolumn{1}{l|}{r-band / uniform weighting} & \multicolumn{1}{l|}{g-band / uniform weighting} \\ \hline
Population 1 (214 detections) & {\color[HTML]{333333} 3.79}         & {\color[HTML]{333333} 4.30}                    & {\color[HTML]{333333} 1.14}                    \\ \hline
Population 2 (224 detections) & {\color[HTML]{333333} 10.4}        & {\color[HTML]{333333} 4.82}                    & {\color[HTML]{333333} 0.46}                    \\ \hline
Population 3 (200 detections) & {\color[HTML]{333333} 2.95}         & {\color[HTML]{333333} 0.42}                    & {\color[HTML]{333333} 0.14}                    \\ \hline
\end{tabular}
\caption{This is the Bayes factor results of Population 1, 2, 3 among r-band weighting, g-band weighting, and uniform weighting models. They are 3 5000-BBH populations with 214, 224, and 200 detections respectively. The simple Powerlaw rate evolution model parameter $\gamma$ was fixed to be 0, i.e. no rate evolution. }
\label{tab:200fix}
\end{table}

\begin{figure}[htbp] 
    \centering
    \includegraphics[width=0.9\textwidth]{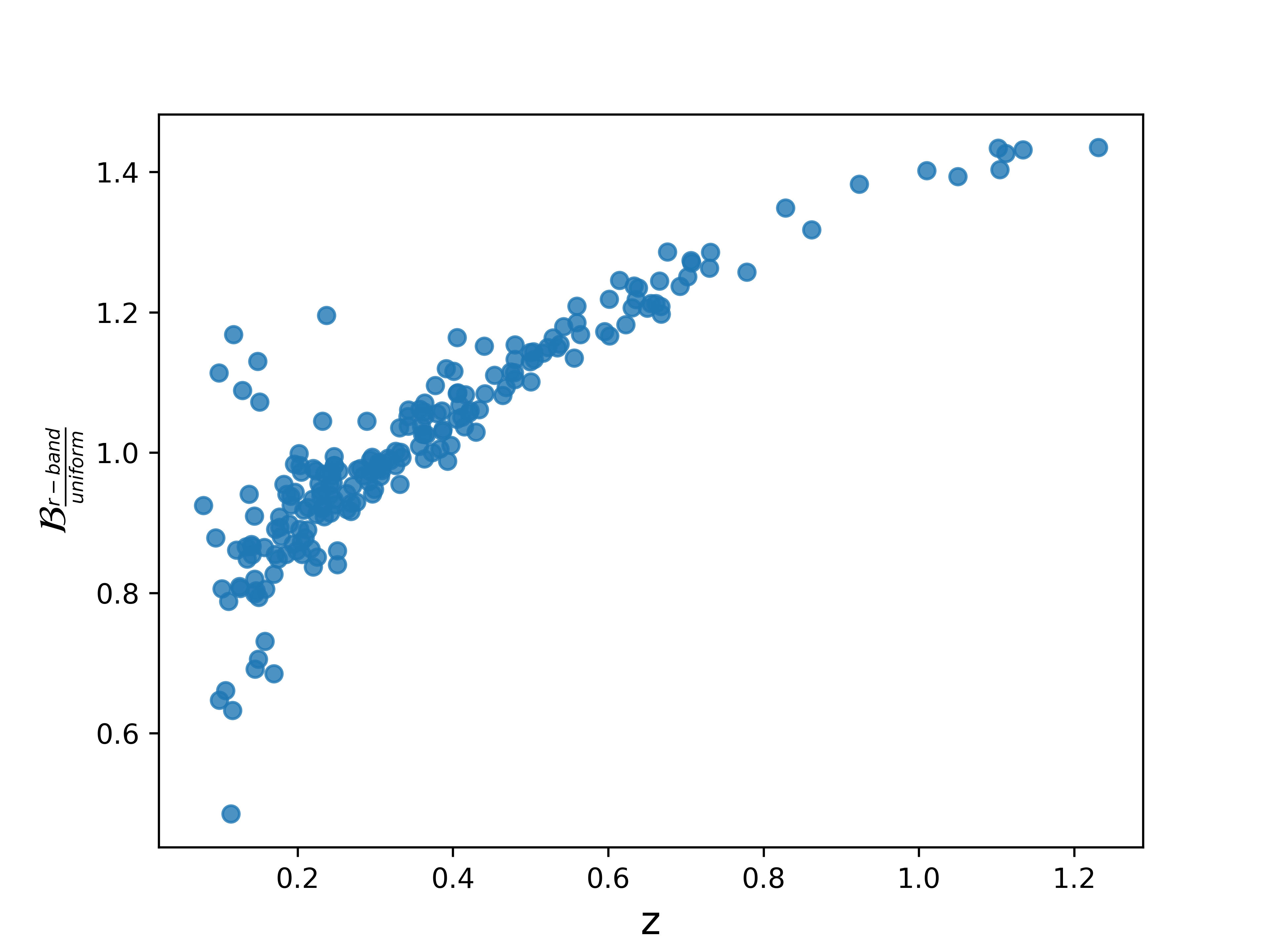}
    \caption{These are r-band weighting versus uniform weighting Bayes factor results, $\mathcal{B}_{\frac{r-band}{uniform}}$, plotted against redshift for individual events in Population 1, with $\gamma$ fixed to be 0. It is clear that there is correlation between the redshift of the event and the Bayes factor.}
    \label{fig:individual_runi}
\end{figure}

Tab.\ref{tab:200fix} illustrates the Bayes factors between the different host galaxy weighting models with r-band weighting being the truth. However, the results of the three populations are not entirely consistent. For example, for Population 1 and Population 2, r-band is favoured against uniform weighting at a Bayes factor $\mathcal{B}=4.30,4.82$ but for Population 3 r-band is slightly disfavoured ($\mathcal{B}=0.42$). Although r-band weighting is closer to g-band weighting than uniform weighting, the Bayes factors for r-band over g-band are still smaller than the Bayes factors for r-band over g-band for Population 2 and 3. To explain the reason for this issue, we plot the distribution of Bayes factors of r-band weighting versus uniform weighting, $\mathcal{B}_{\frac{r-band}{uniform}}$, for individual detections in Population 1 against redshift in Fig.\ref{fig:individual_runi}. According to Fig.\ref{fig:individual_runi}, the Bayes factors for individual detections are correlated with their redshift, and thus the overall Bayes factors should fluctuate with the redshift distribution of detected events. This redshift evolution of Bayes factors is the result of the \ac{bbh} rate evolution introduced by host galaxy weighting models. Different host galaxy weighting models will assume different \ac{bbh} rate evolution, so the redshift of individual events will have an effect on their Bayes factors. 

\begin{figure}[htbp] 
    \centering
    \includegraphics[width=0.9\textwidth]{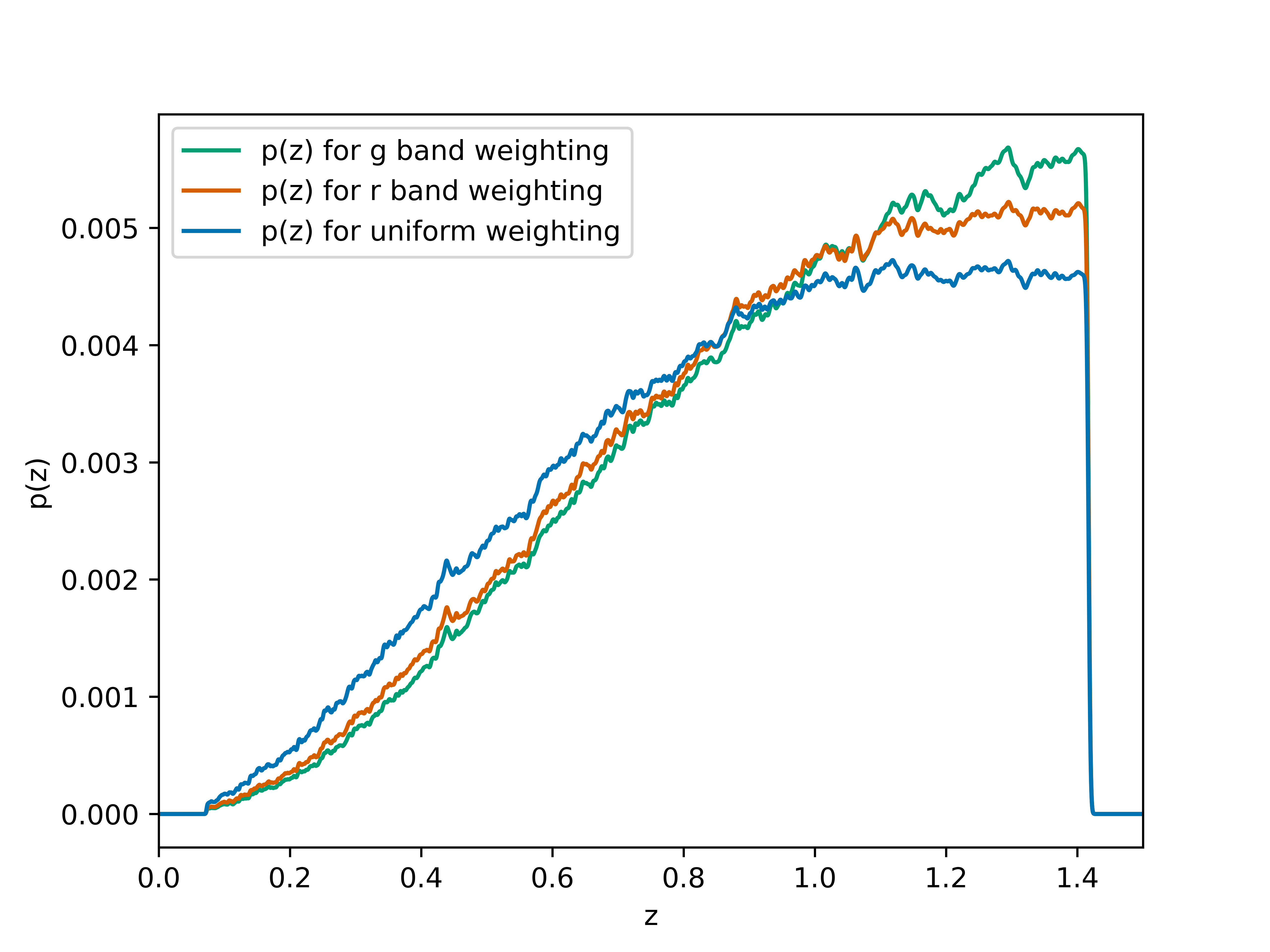}
    \caption{This figure illustrates the overall full-sky redshift priors of MICECATv2 for r-band weighting, g-band weighting, and uniform weighting. It shows that different host galaxy weighting models have different rate evolution assumptions, e.g. uniform weighting prefers low redshift events than r-band weighting.}
    \label{fig:rate_evo}
\end{figure}

The rate evolution assumption of each host galaxy weighting model is shown in Fig.\ref{fig:rate_evo} as the full-sky redshift prior. 
According to Fig.\ref{fig:rate_evo}, if the uniform weighting is the truth, there will be more events at low redshift relative to high redshift ones compared to r-band weighting, thus more low redshift events will produce a Bayes factor that favours the uniform weighting model to the r-band luminosity weighting model. However, since the true model we used for injecting \ac{bbh}s is the r-band weighting, the \ac{bbh} populations follow the rate evolution of full-sky r-band weighted redshift prior. This is why when \ac{los} redshift priors of other weighting models are used, it is necessary to apply a separate rate evolution model to compensate for the difference between the full-sky redshift prior of the model in use and that of the true model. This is also mentioned in \citet{gwcosmo1} which explains that the rate evolution model in the script is meant to represent any extra redshift evolution in addition to the full-sky \ac{los} redshift priors.

It should also be noted that although in Fig.\ref{fig:individual_runi} the redshift strongly correlates with Bayes factors, this correlation is weaker at low redshift. The low redshift events have more random Bayes factors compared to high redshift events, this is because low-redshift events are likely to be better localised so the information from localisation of individual events actually becomes more important. This will be discussed in detail with Fig.\ref{fig:pop45_individual} later.

\subsection{Rate evolution model parameter $\gamma$ unfixed case}\label{Sec:4.2}
\begin{figure}[htbp] 
    \centering
    \includegraphics[width=0.9\textwidth]{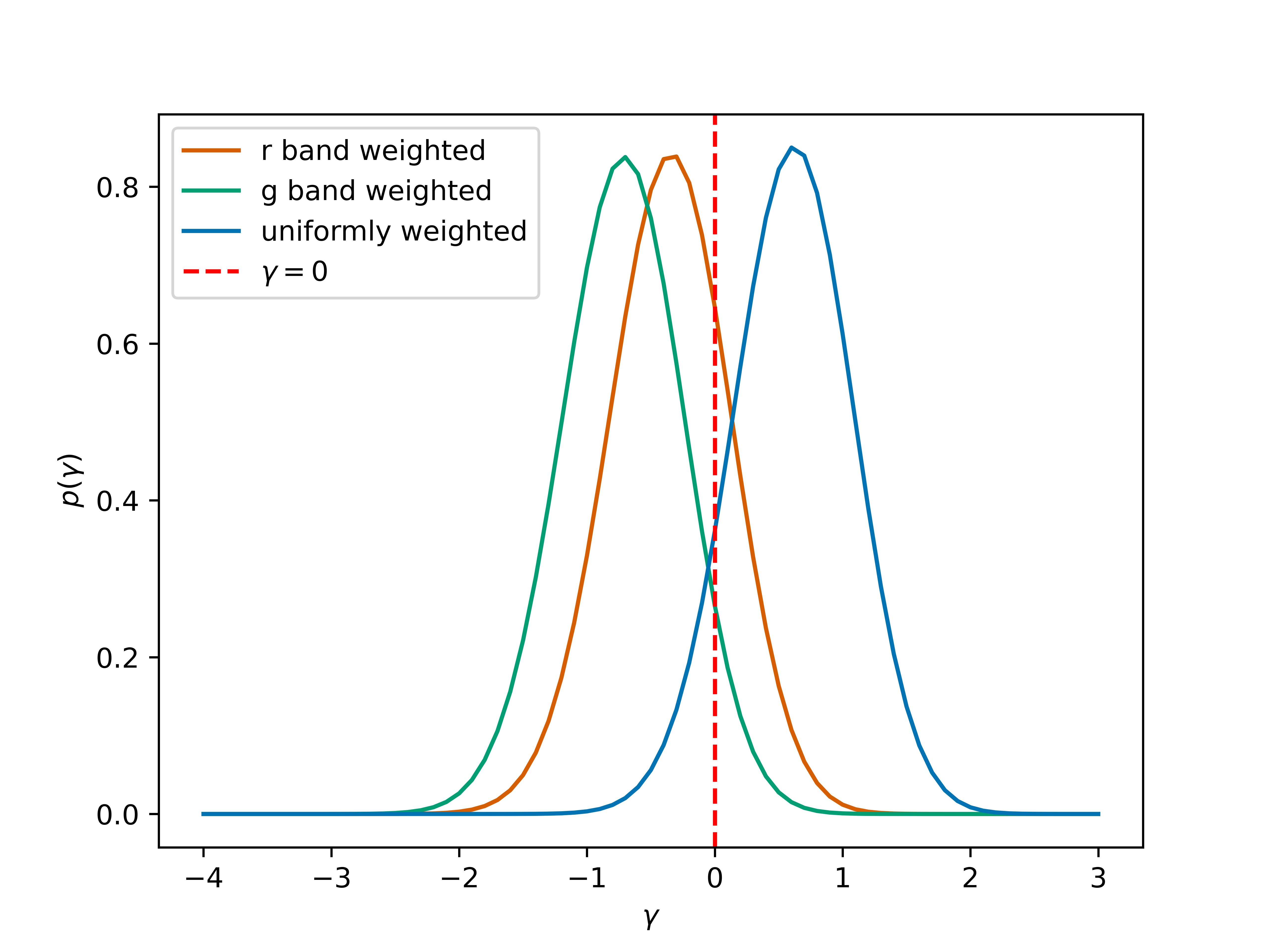}
    \caption{This is the posterior of $\gamma$ for Population 1 for the three host galaxy weighting models. It is clear that all three results do not peak at 0 and do not peak at the same place, implying that they are compensated differently and contain different rate evolution assumptions.}
    \label{fig:gamma_post}
\end{figure}

\begin{figure}[htbp] 
    \centering
    \includegraphics[width=0.9\textwidth]{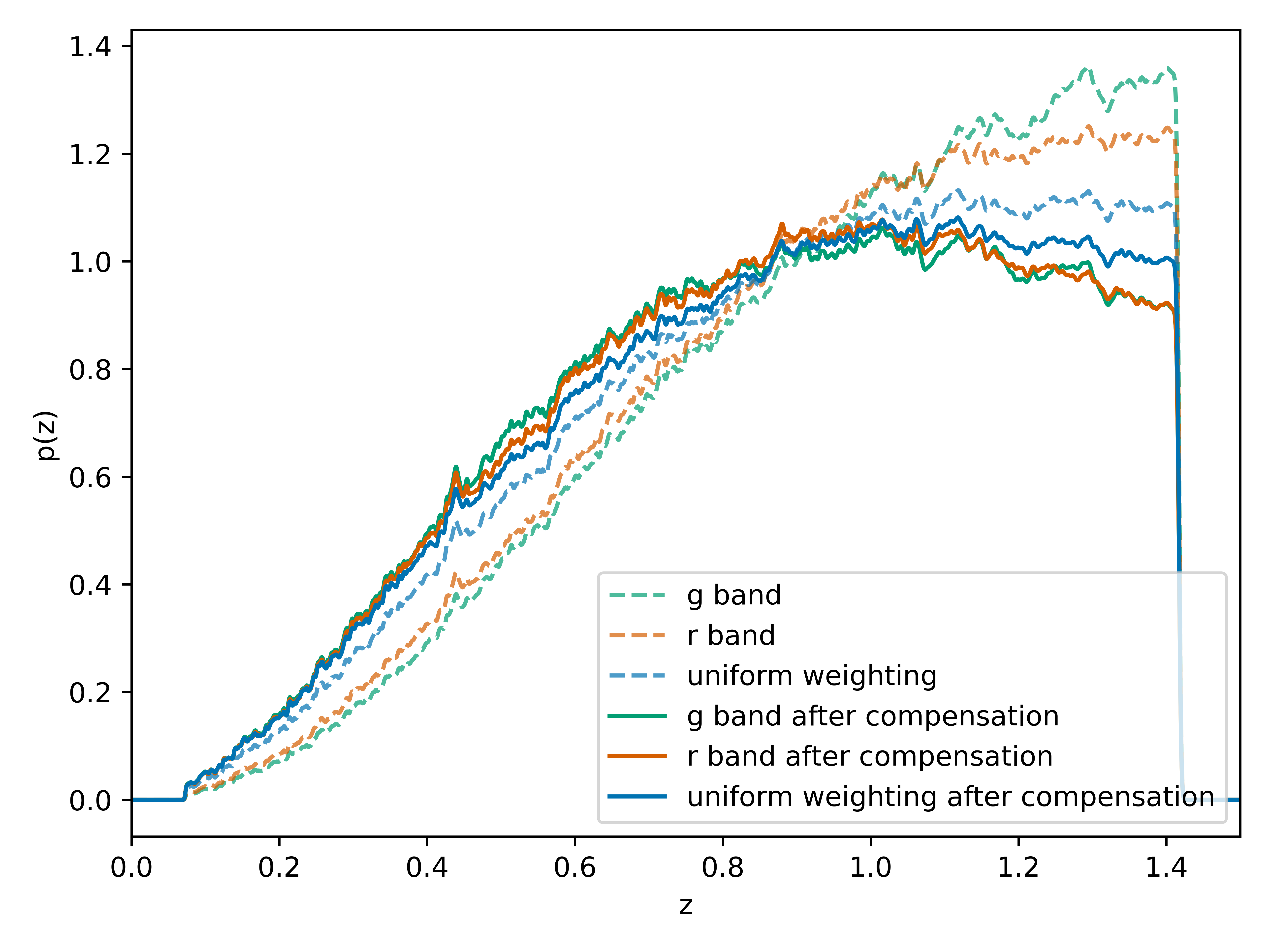}
    \caption{This figure compares the original overall full-sky redshift priors in Fig.\ref{fig:rate_evo} (dashed lines) and the compensated full-sky redshift priors for all three host galaxy weighting models (solid lines). It shows how the compensation pushes the three full-sky redshift priors towards each other, thus compensating the information from the \ac{bbh} rate evolution contributing to the Bayes factors, making the Bayes factors depend more on localisations of individual events. }
    \label{fig:pz_recon}
\end{figure}

    We also carry out tests with the rate evolution model in Eq.\ref{Eq:evidence} unfixed and then marginalised over. This will compensate for these differences in redshift evolution among different host galaxy weighting models. In this test, we apply a simple PowerLaw model, Eq.\ref{Eq:powerlaw}, to the same populations in Tab.\ref{tab:200fix} with parameter $\gamma$, the Powerlaw index, being unfixed. The posteriors of $\gamma$ for the Population 1 in Tab.\ref{tab:200fix} are shown in Fig.\ref{fig:gamma_post} and it is clear that the Powerlaw rate evolution model compensated the three host galaxy weighting models differently. Furthermore, we reconstruct the compensated full-sky redshift prior for the three host galaxy weighting models by multiply the estimated rate evolution model with full-sky redshift priors: $p(z)p(s|z)$. We then compare it with the original full-sky redshift prior in Fig.\ref{fig:pz_recon}. From Fig.\ref{fig:pz_recon} we can see that the compensation, i.e. marginalizing over the \ac{bbh} rate evolution model parameter $\gamma$, did try to ``push'' the full-sky redshift priors of the three models closer to each other. Although it is still not perfectly aligned, there is no doubt that the rate evolution has less effect on the Bayes factors.

The Bayes factor results for the whole three populations are shown in Tab.\ref{tab:200unfix}.
\begin{table}[]
\begin{tabular}{|l|c|c|c|}
\hline
Bayes factors                 & \multicolumn{1}{l|}{r-band / g-band} & \multicolumn{1}{l|}{r-band / uniform weighting} & \multicolumn{1}{l|}{g-band / uniform weighting} \\ \hline
Population 1 (214 detections) & {\color[HTML]{333333} 1.56}         & {\color[HTML]{333333} 2.40}                    & {\color[HTML]{333333} 1.54}                    \\ \hline
Population 2 (224 detections) & {\color[HTML]{333333} 1.99}        & {\color[HTML]{333333} 17.6}                    & {\color[HTML]{333333} 8.87}                    \\ \hline
Population 3 (200 detections) & {\color[HTML]{333333} 0.60}         & {\color[HTML]{333333} 1.77}                    & {\color[HTML]{333333} 2.93}                    \\ \hline
\end{tabular}
\caption{This is the Bayes factor results of the same Population 1, 2, 3 in Tab.\ref{tab:200fix} among r-band weighting, g-band weighting, and uniform weighting models. The simple Powerlaw rate evolution model parameter $\gamma$ was unfixed and then marginalised to calculate the Bayes factors.}
\label{tab:200unfix}
\end{table}
Here Tab.\ref{tab:200unfix} show a result that is more consistent with Fig.\ref{fig:rate_evo} than Tab.\ref{tab:200fix}. The Bayes factors distinguish r-band weighting and uniform weighting better than r-band and g-band with bigger $|\log(\mathcal{B})|$, which is as expected from Fig.\ref{fig:rate_evo} since r-band weighting is more different from uniform weighting than g-band weighting. 

To show the effect of having more \ac{gw} events, we conduct another test with two $\sim$1000~detection populations (Population 4 and 5). The result is shown in Tab.\ref{tab:1000unfix}. The first two rows show the Bayes factor results of Population 4 and 5 while the the last two rows show the Bayes factors of Population 4 without the 10 most informative events in each case (the informativeness of each individual event is evaluated by its own Bayes factor results).
\begin{table}[]
\begin{tabular}{|l|c|c|}
\hline
Bayes factors                  & \multicolumn{1}{l|}{r-band / g-band} & \multicolumn{1}{l|}{r-band / uniform weighting} \\ \hline
Population 4 (1151 detections) & {\color[HTML]{333333} 8.61}        & {\color[HTML]{333333} $9.75\times10^4$}                \\ \hline
Population 5 (1062 detections) & {\color[HTML]{333333} 1.76}         & {\color[HTML]{333333} $1.55\times10^3$}                 \\ \hline
Population 4 without top 10 $\mathcal{B}_{\frac{r-band}{g-band}}$ events & {\color[HTML]{333333} 2.04}        & {\color[HTML]{333333}\diagbox{}{} }                \\ \hline
Population 4 without top 10 $\mathcal{B}_{\frac{r-band}{uniform}}$ events & {\color[HTML]{333333}\diagbox{}{} }        & {\color[HTML]{333333} $7.22\times10^3$}                \\ \hline
\end{tabular}
\caption{The first two rows are the Bayes factor results of Population 4 and 5 among r-band weighting, g-band weighting, and uniform weighting models. They are 2 25000-BBH populations with 1151 and 1062 detections respectively. The last two rows are the Bayes factor results for Population 4 excluding the 10 most informative events in each case. To be more specific, we calculated the Bayes factor results $\mathcal{B}_{\frac{i}{j}}$ of each individual events, and excluded the events with 10 highest Bayes factors in the according case. The \ac{bbh} rate evolution parameter $\gamma$ was unfixed and later marginalised.}
\label{tab:1000unfix}
\end{table}
From the first two rows of Tab.\ref{tab:1000unfix}, it is clear that the Bayes factors between r-band weighting and uniform weighting increase significantly with more \ac{bbh} events compared to Tab.\ref{tab:200unfix}. There is also improvement on r-band versus g-band especially in Population 4. This means the method distinguishes these host galaxy weighting models better with more \ac{bbh} detections, especially for host galaxy weighting models that are more different, e.g. r-band and uniform weighting here. 

Another point in Tab.\ref{tab:1000unfix} is although Population 4 and 5 have similar numbers of events, the Bayes factors of Population 4 is bigger than those of Population 5 for around a magnitude. However, by excluding ten 10 most informative events for both cases (third and fourth row), The Bayes factors drop to the same magnitude of Population 5 (second row). This means the full Population 4 results (first row) are dominated by a small number of highly informative events. This implies that very well-localised \ac{cbc} mergers, also known as gold dark sirens, can be very important in distinguishing host galaxy weighting models.

\begin{figure}[htbp] 
    \centering
    \includegraphics[width=0.9\textwidth]{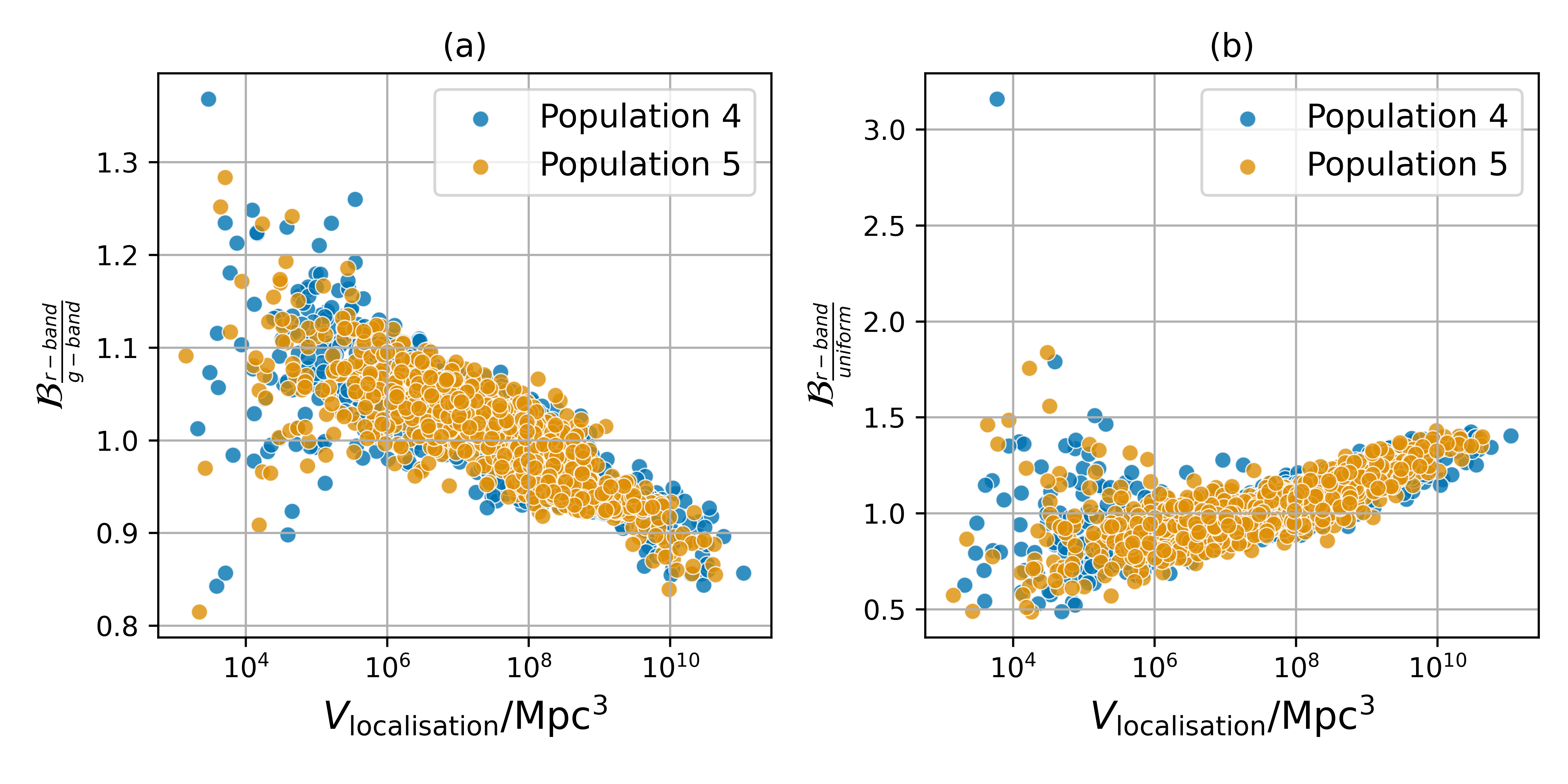}
    \caption{This figure show the Bayes factors among the three host galaxy weighting models marginalised over the $\gamma$ posteriors of individual events in Population 4 and 5. The left panel (a) is for Bayes factors of r-band versus g-band while the right panel (b) is Bayes factors for r-band versus uniform weighting. The x-axis represents the localisation volume of the \ac{bbh} events within 99\% of the sky localisation area and $d_L\pm3\sigma_{d_L}$.}
    \label{fig:pop45_individual}
\end{figure}

To further illustrate this point, we show the distribution of Bayes factors plotted against the localisation volume for each event. The volume covers 99\% of the sky localisation area and $d_L\pm3\sigma_{d_L}$. The figure is shown in Fig.\ref{fig:pop45_individual}. First of all we can see a clear correlation between the localisation volume, $V_{\mathrm{localisation}}$, and Bayes factors, although different correlations for different Bayes factors. This is still because of the rate evolution model (as $V_{\mathrm{localisation}}$ has a strong correlation with $z$) but this does not mean our overall results are dominated by rate evolution information and marginalisation over $\gamma$ does not work. In fact, the marginalisation over $\gamma$ method only works on population level, not on individual event level since one event contributes little to the $\gamma$ posterior. The high redshift events and low redshift events will balance the rate evolution information of each other. Second, at low $V_{\mathrm{localisation}}$, i.e. when events are really well-localised, the individual event Bayes factors deviate from the trend, meaning that it is driven by specific galaxy information within the localisation, rather than the overall redshift distribution. At low $V_{\mathrm{localisation}}$, while r-band versus g-band Bayes factors in the left panel seem rather random, the r-band versus uniform weighting Bayes factors in the right panel clearly go against the trend and favour r-band. This is expected since uniform weighting is much more different to r-band than g-band. The figure also shows the significant contribution from a small number of well-localised events, for example the most-informative event in the right panel has a Bayes factor over 3.

\subsection{Madau-Dickinson rate evolution model unfixed case}\label{Sec:4.3}
We also conduct a test to use Madau-Dickinson rate evolution model to find out if a more complicated and flexible rate evolution model compensates the rate evolution better than the simple Powerlaw model. The mathematical expression of Madau-Dickinson model is given in Eq.\ref{Eq:madau}. we have unfixed $\gamma$, $k$, and $z_p$ and marginalised over the parameter space using a Bayesian inference tool BILBY and a nested sampling tool NESSAI in our test \citep{bilby,nessai}.

Given the computational cost, we only test Madau-Dickinson model with Population 1 mentioned in Tab.\ref{tab:200fix} and Tab.\ref{tab:200unfix}. The result is shown in Tab.\ref{tab:md1}, which compares the result using Madau-Dickinson model with that using the simple Powerlaw model. However, the difference is trivial, possibly implying a more sophisticated rate evolution model is needed here or the inherent structural differences among the full-sky redshift priors cannot be perfectly compensated using Powerlaw-like models in MICECATv2.
\begin{table}[]
\begin{tabular}{|l|c|c|}
\hline
Bayes factors for Population 1 (214 detections)& \multicolumn{1}{l|}{r-band / g-band} & \multicolumn{1}{l|}{r-band / uniform weighting} \\ \hline
 Madau-Dickinson model & {\color[HTML]{333333} 1.66}        & {\color[HTML]{333333} 2.26}  \\ \hline
  Simple Powerlaw model & {\color[HTML]{333333} 1.56}        & {\color[HTML]{333333} 2.40}
\\ \hline
\end{tabular}
\caption{This is the comparison of Bayes factors for Population 1 using Madau-Dickinson rate evolution model and simple Powerlaw model. The simple Powerlaw result is taken from Tab.\ref{tab:200unfix}. The Madau-Dickinson rate evolution parameters $\gamma$, $k$, and $z_p$ were unfixed and later marginalised.}
\label{tab:md1}
\end{table}


\section{Conclusion}\label{Sec:5}
In this paper, we test the possibility of distinguishing \ac{bbh} host galaxy weighting models using gwcosmo with a mock galaxy catalogue drawn from MICECATv2 and mock \ac{bbh} \ac{pe} results from GWFish. Our method calculates the Bayes factors between different host galaxy weighting models by comparing their evidence in the standard gwcosmo dark siren cosmology analysis. We assume fixed cosmology and \ac{bbh} population during the analysis. The analysis is conducted with r-band (``true'' model in our test) and g-band luminosity weighting models as well as uniform weighting under an \ac{o5} sensitivity three detector network including LIGO-Hanford, LIGO-Livingston, and Virgo. The three mock populations have around 200 detections each. We should note that by reducing the number of galaxies for computational reasons, our results will be more optimistic. However the choice of two close bands (DESI r-band and g-band) balance our r-band versus g-band Bayes factor results with pessimism. In addition to this choice of close bands, the fact that we do not consider KAGRA in this paper also makes our results more pessimistic \citep{kagra}. In future papers we would like to consider KAGRA as well but that requires the resolution of \ac{los} redshift priors to be even higher.

We calculate Bayes factors for both cases where the rate evolution model is fixed and unfixed. For the rate evolution fixed case, we find that Bayes factor results are not consistent. This is because the rate evolution of \ac{bbh}s will add information in addition to the information from the localisation of individual events. For the rate evolution model unfixed case, the simple Powerlaw rate evolution model parameter $\gamma$ is unfixed and then marginalised to compensate the information from rate evolution. This means the Bayes factors depend more on the localisation of individual \ac{gw} events. The results align with our expectation: an overall slight preference of r-band over g-band weighting and a strong preference of r-band over uniform weighting, as predicted from the differences among the full-sky redshift priors of these models. 

We also test the case with the number of \ac{bbh} events increased to around 1000 detections per population. As expected, increasing event numbers do make the Bayes factors more substantial with rate evolution model unfixed, especially for host galaxy weighting models that are less similar. For both $\sim$1000-detection cases, there is decisive support for r-band over uniform weighting. It should also be noted that a small number of well-localised events can dominate the result, highlighting the importance of golden dark sirens.

Besides the simple Powerlaw rate evolution model, the more sophisticated Madau-Dickinson rate evolution model is also tested to improve the compensation. However, the change in Bayes factors is small.

We show that for future \ac{o5}/A\# sensitivity network and improved galaxy catalogue, it is possible to distinguish different weighting models with strong evidence (Bayes factors over 10) under the assumption of fixed cosmology, fixed \ac{bbh} population, and unfixed rate evolution model. This paper also highlights that in the current dark siren cosmology analysis, the host galaxy weighting and the rate evolution estimation are correlated, and the rate evolution estimation is only part of the actual rate evolution on top of what is already accounted by the full-sky redshift priors of host galaxy weighting models, as was already explained in \citet{gwcosmo1}.

For future investigations, it would be useful to add more reality about the galaxy catalogues such as accounting for the incompleteness and more realistic redshift uncertainties, the impact of faint galaxies will also need to be investigated, and using a more complicated rate evolution model other than Madau-Dickinson could improve the results as well, etc. More realistic tests on different host galaxy weighting models will help us to distinguish them in the future with \ac{o5}/A\# or even third-generation detectors, and will also be important for BBH formation channel as well as \ac{bbh} rate evolution studies. 

\section{Acknowledgements}
This paper is funded by the Chinese Scholarship Council. This paper has used computing resources of LIGO-Virgo Collaboration. This work has made use of CosmoHub, developed by PIC (maintained by IFAE and CIEMAT) in collaboration with ICE-CSIC. It received funding from the Spanish government (grant EQC2021-007479-P funded by MCIN/AEI/10.13039/501100011033), the EU NextGeneration/PRTR (PRTR-C17.I1), and the Generalitat de Catalunya.


\bibliography{sample631}{}
\bibliographystyle{aasjournal}

\appendix
\section{Effect of including fainter galaxies}\label{Sec:appendix}
\begin{figure}[htbp] 
    \centering
    \includegraphics[width=0.9\textwidth]{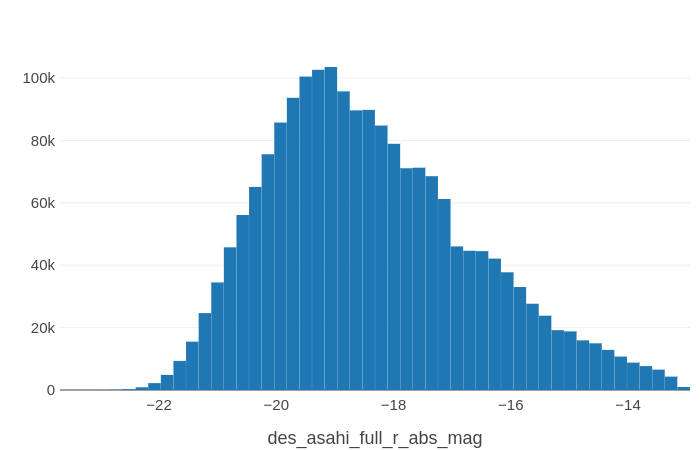}
    \caption{This is the distribution of galaxies on r-band absolute magnitude of MICECATv2. The distribution peak at around -20 because of a lack of faint galaxies at the higher end of the redshift range.}
    \label{fig:micecat}
\end{figure}

For our main results, we choose to cut MICECATv2 at r-band absolute magnitude -20. Here we perform a comparison test that uses a mock catalogue by cutting MICECATv2 at r-band absolute magnitude -19 to test the effect of including fainter galaxies on our results. Here we define this cut as r-band absolute magnitude threshold $M_r^{{th}}=-19$. The distribution of galaxies on r-band absolute magnitude of MICECATv2 is shown in Fig.\ref{fig:micecat} which is generated by COSMOHUB \citep{cosmodc2,cosmohub1}. The distribution in Fig.\ref{fig:micecat} peaks at around -19, meaning there are fainter galaxies missing at higher end of the redshift range. Thus, pushing $M_r^{{th}}$ any further to the fainter end will introduce a serious under-density of faint galaxies at high redshift.

After the luminosity cut, we randomly draw 10\%, the same percentage with main results, from galaxies that passed this threshold and assume that this sliced catalogue is complete. There are around 20.5 million galaxies used, more than doubled compared to our main results which includes 9 million galaxies in total. This implies over half of the 20.5 million galaxies should be faint galaxies with $-19<M_r<-20$. 

\begin{figure}[htbp] 
    \centering
    \includegraphics[width=0.9\textwidth]{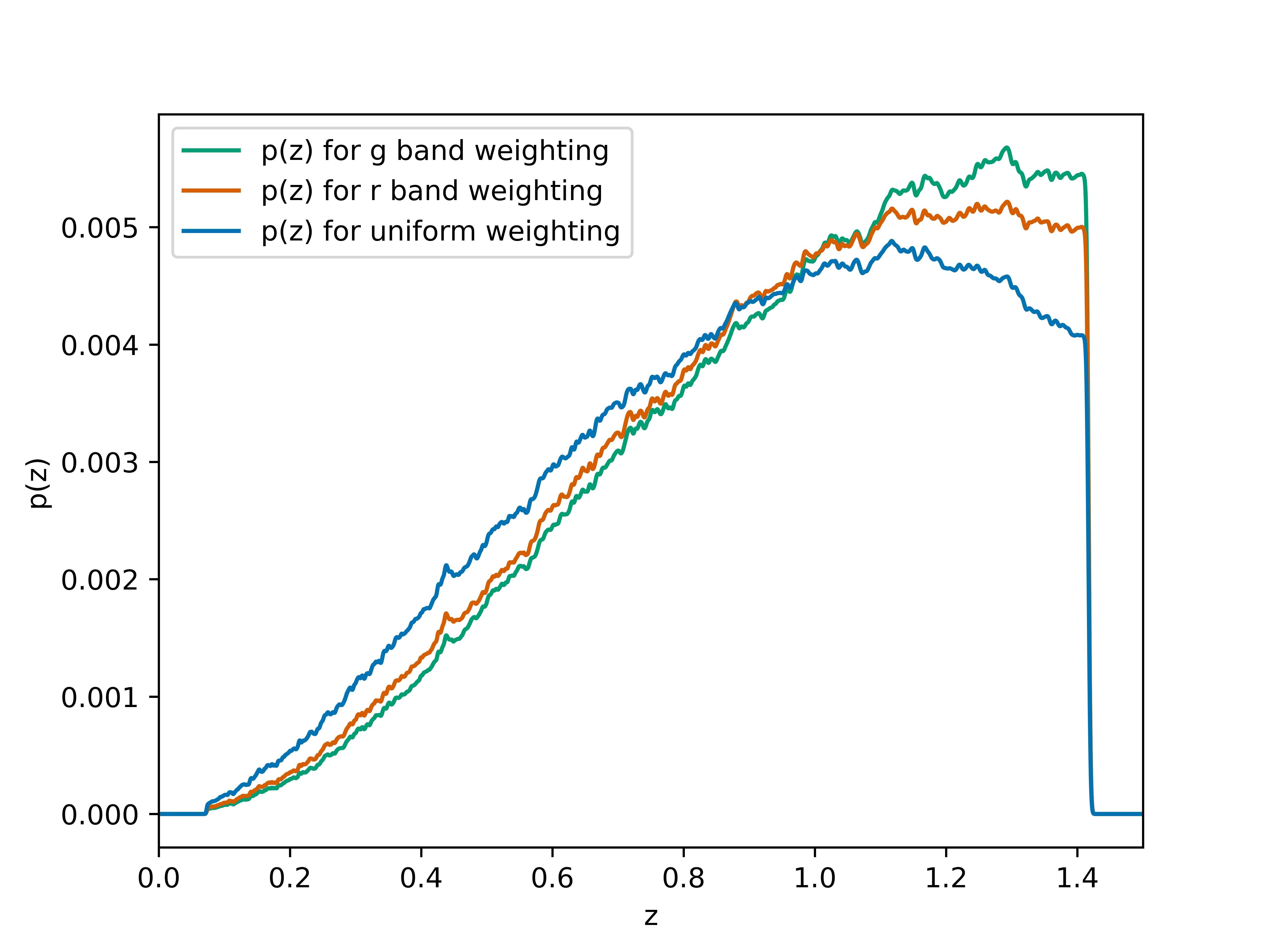}
    \caption{This is the full-sky redshift priors of MICECATv2 for r-band weighting, g-band weighting, and uniform weighting when the r-band absolute magnitude cut $M_r^{{th}}=-19$. The uniform weighted prior goes down after redshift 1.1 while the r-band and g-band weighted priors do not have significant changes compared to Fig.\ref{fig:rate_evo}.}
    \label{fig:pz_all_19}
\end{figure}

\begin{figure}[htbp!] 
    \centering
    \includegraphics[width=0.9\textwidth]{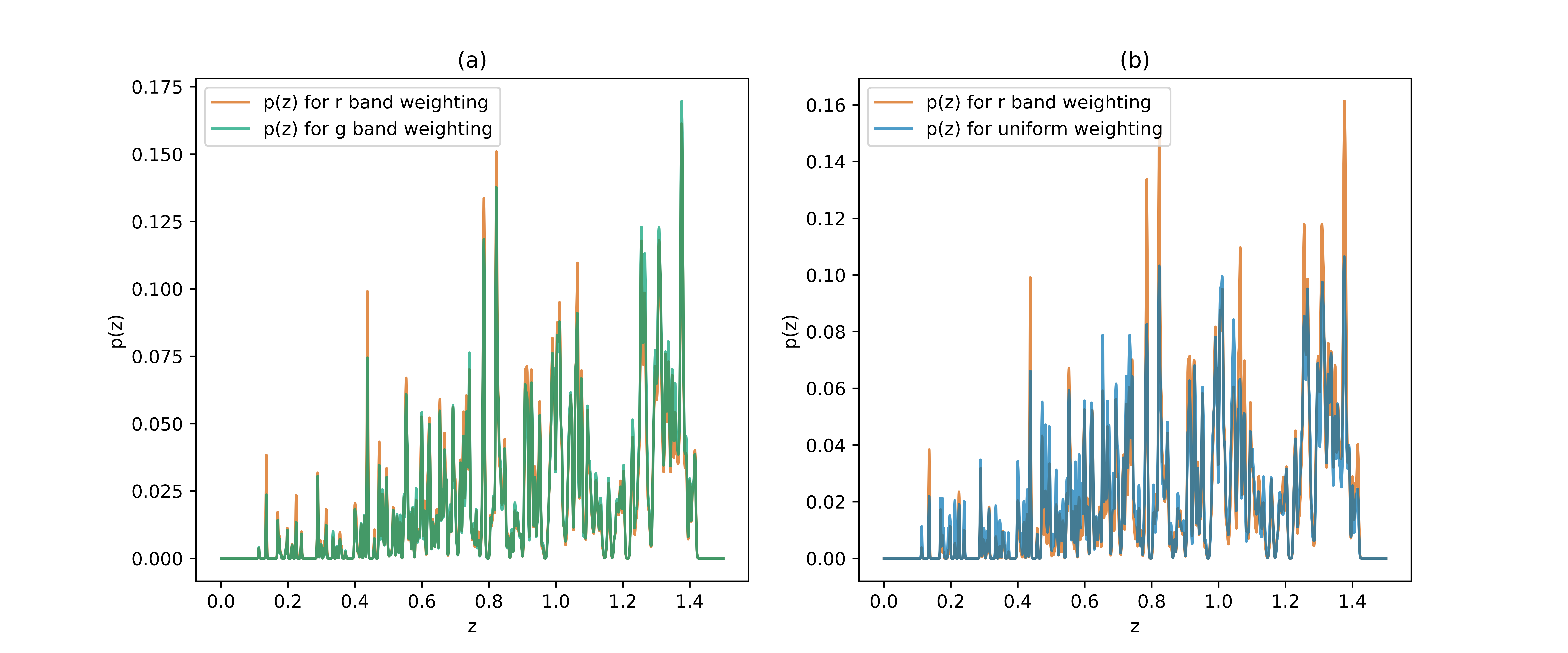}
    \caption{These are the \ac{los} redshift priors at the same pixel in Fig.\ref{fig:example_pixel_runi}. The left panel (a) compares r-band weighting versus g-band weighting while the right panel (b) compares r-band weighting and uniform weighting. There are more fainter galaxies included, as a results the differences among the host galaxy weighting models increased compared to Fig.\ref{fig:example_pixel_runi}.}
    \label{fig:eg_pic_19}
\end{figure}

The full sky redshift prior of this catalogue is shown in Fig.\ref{fig:pz_all_19} and example \ac{los} of a pixel is shown in Fig.\ref{fig:eg_pic_19}. Fig.\ref{fig:pz_all_19} shows that at the uniform weighted redshift prior peaks at around redshift 1.1, which means there is already slight under-density of faint galaxies at high redshift at this $M_r^{{th}}$. However, both r-band and g-band-weighted redshift priors do not change significantly compared to Fig.\ref{fig:rate_evo} because the faint galaxies are down-weighted. Fig.\ref{fig:eg_pic_19} shows the \ac{los} prior of the same pixel shown in Fig.\ref{fig:example_pixel_runi}. We can roughly see that the difference between r-band weighting and uniform weighting is greater in Fig.\ref{fig:eg_pic_19} than that in Fig.\ref{fig:example_pixel_runi}. This is also because the faint galaxies added in the $M_r^{{th}}=-19$ are down weighted by luminosity weighting while having the same weight with other galaxies in uniform weighting. Thus, we expect that by including fainter galaxies it would be easier to distinguish between the luminosity weighting models and the uniform weighting model.

We then generate a $\sim$1000 detection \ac{bbh} population, Population 6, with the same population model with the main results, injected into this $M_r^{{th}}=-19$ catalogue based on r-band luminosity weighting. We then apply our host galaxy weighting model comparison analysis using Population 6. 

\begin{table}[htbp!]
\begin{tabular}{|l|c|c|}
\hline
Bayes factors                  & \multicolumn{1}{l|}{r-band / g-band} & \multicolumn{1}{l|}{r-band / uniform weighting} \\ \hline
Population 4 (1151 detections), $M_r^{{th}}=-20$ & {\color[HTML]{333333} 8.61}        & {\color[HTML]{333333} $9.75\times10^4$}                \\ \hline
Population 5 (1062 detections), $M_r^{{th}}=-20$ & {\color[HTML]{333333} 1.76}         & {\color[HTML]{333333} $1.55\times10^3$}                 \\ \hline
Population 6 (1053 detections), $M_r^{{th}}=-19$ & {\color[HTML]{333333} 3.80}        & {\color[HTML]{333333} $2.59\times10^6$ }                \\ \hline

\end{tabular}
\caption{This table compares the Bayes factor results between Population 6 which is injected into $M_r^{{th}}=-19$ catalogue, and Population 4 and 5 which are injected into $M_r^{{th}}=-20$ catalogue. The Bayes factor of r-band versus uniform weighting for Population 6 significantly increased by $2\sim3$ magnitude compared to Population 4 and 5. }
\label{tab:1000unfix19}
\end{table}

The results of Population 6 compared to Population 4 and 5 are shown in Tab.\ref{tab:1000unfix19}. There is a significant improvement in the Bayes factor comparing r-band and uniform weighting for Population 6 as expected, which increases by $2\sim3$ magnitudes. As mentioned above, this is the result of the fact that fainter galaxies have the same weights with brighter galaxies in uniform weighting but are down weighted by r-band/g-band luminosity weighting. The Bayes factor of r-band over g-band for Population 6 is roughly in the same magnitude with Population 4 and 5, which is expected since the low luminosity galaxies are down weighted by r-band and g-band weighting.

Overall, we find that including fainter galaxies in our analysis will improve our Bayes factor model comparison analysis for luminosity weighting versus uniform weighting. However, we still choose $M_r^{{th}}=-20$ for our main results because it is volume limited, i.e. no under-density of faint galaxies at high redshift.

\end{document}